\documentclass[twocolumn,showpacs,preprintnumbers,amsmath,amssymb, groupedadress, altaffilletter]{revtex4}

\usepackage{epsfig}
\usepackage{amsmath}
\usepackage{graphicx}
\usepackage{bm}
\usepackage{ulem}

\begin{document}

\title{A high-reflectivity high-Q micromechanical Bragg-mirror}

\author{H. R. B\"ohm}
\author{S. Gigan}
 \affiliation{Faculty of Physics,
University of Vienna, Boltzmanngasse 5, A-1090 Vienna, Austria}
\affiliation{Institute for Quantum Optics and Quantum Information
(IQOQI), Austrian Academy of Sciences, Boltzmanngasse 3, A-1090
Vienna, Austria}

\author{G. Langer}
\affiliation{Institute for Applied Physics,
Johannes-Kepler-University Linz, Altenbergerstr. 69, A-4040 Linz,
Austria}

\author{J. B. Hertzberg}
\affiliation{Department of Physics, University of Maryland, College
Park, MD 20740, USA}

\author{F. Blaser}
\affiliation{Faculty of Physics, University of Vienna,
Boltzmanngasse 5, A-1090 Vienna, Austria} \affiliation{Institute
for Quantum Optics and Quantum Information (IQOQI), Austrian
Academy of Sciences, Boltzmanngasse 3, A-1090 Vienna, Austria}

\author{D. B\"auerle}
\altaffiliation[temporary address:]{Russell S. Springer Professor -
University of California, Berkeley, CA 94720-1740, USA}
\affiliation{Institute for Applied Physics,
Johannes-Kepler-University Linz, Altenbergerstr. 69, A-4040 Linz,
Austria}

\author{K. C. Schwab}
\affiliation{Laboratory for Physical Sciences, University of
Maryland, College Park, MD 20740 USA} \altaffiliation[present
address:]{Cornell University, Ithaca, NY USA}
\author{A. Zeilinger}
\author{M. Aspelmeyer}\email{markus.aspelmeyer@univie.ac.at}
\affiliation{Faculty of Physics, University of Vienna,
Boltzmanngasse 5, A-1090 Vienna, Austria} \affiliation{Institute for
Quantum Optics and Quantum Information (IQOQI), Austrian Academy of
Sciences, Boltzmanngasse 3, A-1090 Vienna, Austria}

\begin{abstract}
We report on the fabrication and characterization of a
micromechanical oscillator consisting only of a free-standing
dielectric Bragg mirror with high optical reflectivity and high
mechanical quality. The fabrication technique is a hybrid 
approach involving laser ablation and dry etching. The mirror has a
reflectivity of 99.6\%, a mass of 400ng, and a mechanical quality
factor Q of approximately $10^4$. Using this micromirror in a Fabry
Perot cavity, a finesse of 500 has been achieved. This is an
important step towards designing tunable high-Q high-finesse
cavities on chip.
\end{abstract}


\clearpage

\maketitle

Micromechanical oscillators are today widely used in applications
from thermal, infrared and chemical to biological sensing
\cite{Sensors}. This huge success is due to the fact that sensors
based on micro cantilevers can detect extremely small stimuli such
as temperature and mass changes as well as small external
forces~\cite{Ekinci_2005}. Available devices can be placed in two
categories based on their readout scheme. Micro electro-mechanical
systems (MEMS) use a wide array of electronic coupling schemes to
transduce mechanical energy into electronic signals, while micro
opto-mechanical systems (MOMS) are usually read out using an optic
lever or a Fabry Perot interferometer. For MOMS, the sensitivity or
coupling strength is mainly dependent on both the mechanical quality
and the reflectivity of the cantilever. By increasing the mechanical
quality factor and using ultra-high reflectivity materials one can
thus considerably increase the performance of such sensors.
Ultimately, when all technical noise sources have been eliminated,
quantum mechanics poses a limit in the sensitivity of such
devices~\cite{WeakForces}. Although in today's sensor applications
this quantum limit has not yet been reached~\cite{Millburn}, a very
close approach has been demonstrated~\cite{Lahaye, Hadjar}.

\begin{figure}[htbp]
\centerline{\includegraphics[width=0.4\textwidth]{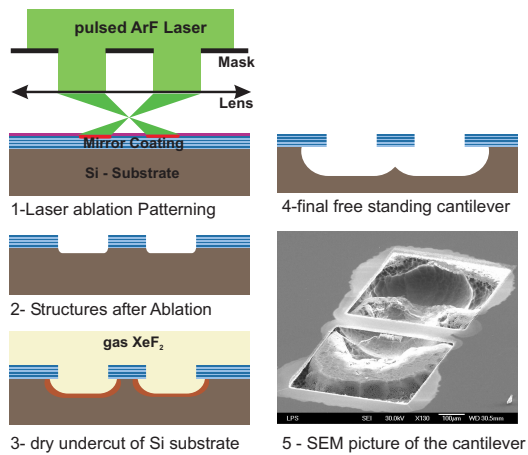}}
\caption{The steps of the fabrication process: (1) The Si-Substrate
coated with a high reflectivity bragg mirror is protected from
debris of the ablation process with a thin layer of photoresist.
Using a simple imaging system the light from a $F_2$ laser ablates
rectangular structures from the coating. The ablation is stopped as
soon as the Si-Layer is reached.(2) The debris from ablation is
removed together with the photoresist in a solvent bath. (3) Using
$XeF_2$ gas the silicon substrate is selectively etched, leading to
an undercut of the beam. (4) The final structure consists only of a
freestanding Bragg mirror. (5) The oscillator is the bridge in the
center. It is about 520 $\mu m$ long, 120 $\mu m$ wide and has a
total thickness of about 2.4 $\mu m$. The dark region of the surface
corresponds to the region where the HR coating has not been
undercut.} \label{fab}
\end{figure}

While the reflectivity of a bulk dielectric or semiconductor
material is, in general, quite low (typically around 50\%),
reflectivities of up to 98\% can be achieved by depositing a thin
metallic layer on top of predefined structures. If higher
reflectivities are needed one has to use Bragg mirrors which consist
of a stack of thin layers of materials with different refractive
indices. Such mirror materials are widely used and can reach
reflectivities larger than 99.999\% or, when used as mirrors in an
optical cavity, a cavity finesse of  $>10^6$. High reflectivities
have been achieved in MEMS tunable Fabry-Perot etalons, however with
 a very low mechanical quality Q typically well below
10~\cite{Tucker}. In this paper we report the fabrication and
characterization of a high-quality mechanical beam oscillator consisting of a free-standing Bragg mirror with $99.6\pm0.1\%$
reflectivity and of high mechanical quality $Q\approx10^4$. In
contrast to previous approaches, where the reflectivity of
Si-microstructures is increased by coating them in a postprocessing
step, we directly fabricate the microstructure out of a large-scale
Bragg-mirror (See figure \ref{fab}). This avoids the unwanted side
effects that typically arise during the coating procedure, such as
bending due to thermal mismatch~\cite{Stressdielectric}. The
fabrication technique is based on pulsed-laser ablation of the
coating, followed by dry-etching of the substrate underneath. Laser
ablation is an interesting technique complementary to standard
micro-fabrication methods, since, unlike wet etching or reactive ion
etching, it does not depend on the chemical reactivity of the
material being patterned \cite{Bauerle, Schafer}. Furthermore, the
good spatial selectivity achieved with short laser pulses allows the
local removal of material while it preserves the coating quality on
the oscillator.

\begin{figure}[htbp]
\centerline{\includegraphics[width=0.4\textwidth]{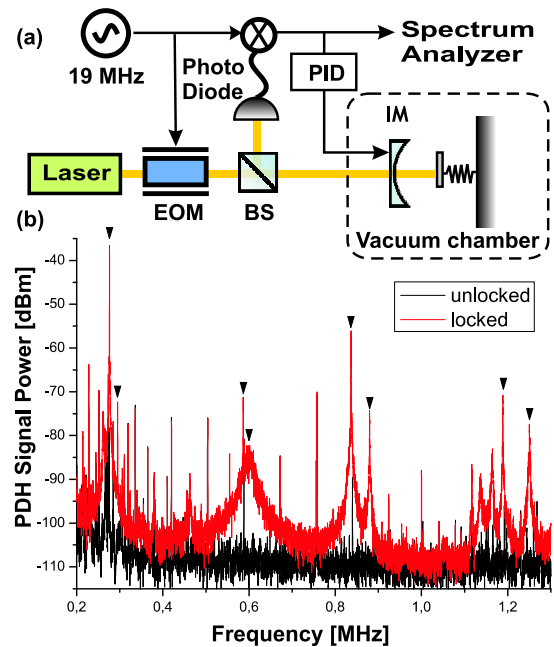}}
\caption{(a)~Sketch of the Measurement scheme. The laser is an
ultrastable YAG Laser which delivers up to 1W continuous-wave laser
light at 1064nm with 1 kHz linewidth. The laser is phase modulated
via an electro-optic modulator (EOM) at 19 MHz. This frequency is
chosen to be above the mechanical resonance frequencies, but within
the cavity bandwidth.
About 0.5 mW of this modulated light is used to pump the optical
cavity. Using a PBS cube, we send part of the light reflected back
from the cavity to a High-Speed InGaAs photodiode.
By mixing the photocurrent with the optical modulation frequency
and subsequent low-pass filtering we derive the Pound-Drever-Hall
(PDH)\cite{pdh} error signal that is, to the first order,
proportional to the cavity length. Feeding this signal to a PID
controller that drives a piezoelectric actuator the length of the
cavity is stabilized at resonance. Due to the limited bandwidth of
the control loop, the cavity length is kept constant at accoustic
frequencies, while above the cut-off frequency of the PID
controller the vibrations of the beam are still present in the
error signal. Therefore feeding the error signal to a spectrum
analyzer allows us to monitor both cavity and mirror dynamics.
(b)~Frequency spectrum of the PDH signal when the cavity is
locked (red) and not locked (black). The trace reflects the
position RMS noise of the cantilever. Markers indicate mechanical
resonance peaks (see text). The noise floor
corresponds to approximately $10^{-16}m/\sqrt{Hz}$.} \label{setup}
\end{figure}

The low selectivity on material at high laser fluences allows
ablation of thick dielectric stacks of materials with different
chemical properties. The good control of the ablation depth permits
well-defined removal of the coating. To free the structure from the
substrate, a very selective dry-etching method was employed. It
undercuts the patterned coating but does not etch it in any
measurable manner. Following the description of the fabrication
method we present the experimental setup to characterize the optical
and mechanical quality of the resulting structure. Subsequently we
give a short summary and discussion of the obtained results.

The fabrication of the micromirrors starts from a standard polished
silicon wafer on which 16 layers of $\text{TiO}_2$ and
$\text{SiO}_2$ have been deposited to form a highly reflective (HR)
Bragg mirror. This process, which has been optimized to reduce
tensile stress, was done by the coating company O.I.B.GmbH. The
nominal reflectivity is 99.8\% at a wavelength of 1064 nm.

Subsequently the coated wafer is patterned using laser ablation by
projecting 193 nm ArF-laser radiation ( 1.3 $J / cm^2 $, $\tau_l =
28$ ns, repetition rate 1 to 10 pps) via a mechanical mask
(reduction optics 10:1). With the laser pulse length employed, the
heat affected zone (HAZ) was about 5 $\mu m$. This is in reasonable
agreement with estimates of lateral heat diffusion in the silicon
substrate \cite{Bauerle}. To protect the mirror from debris
generated during laser ablation, the surface is coated with a thin
layer of a soft-baked photoresist (Shipley S1813). After ablation
this layer is dissolved in an ultrasonic bath of acetone. By putting
the sample in a $XeF_2$ atmosphere the exposed Si substrate is
etched rapidly \cite{Chang} (1 $\mu m$ / min) isotropically and very
selectively in the ablated regions around and below the
beam. The $Si$/$SiO_2$ selectivity of the etch is better than 1000:1
\cite{Manufact}, and there is no measurable etch of TiO$_2$, which
is the top layer of the coating. The etching is optically monitored
in-situ, and is stopped as soon as the beam is fully underetched.
The final structure consists of a free-standing beam of
approximately 520 $\mu m$ length and a width of 120 $\mu m$, which
is made only out of the original 2.4 $\mu m$ thick Bragg-mirror
coating surrounded by a membrane of about half of its width. The
total mass of the beam, as calculated from its dimensions, is about
400ng. The uneven undercut of the structure depicted in figure
\ref{fab} is most probably due to partial coverage of the silicon
surface with other materials. Possible sources of contamination are
incomplete ablation, redeposition of ablated material or oxidation
of the heated silicon surface in air, all of which can be avoided by
ablating the coating in vacuum or deeper ablation.

The micro-mirror was characterized via Fabry-Perot interferometry.
We built a linear Fabry-Perot optical cavity with the beam as the
highly-reflecting end mirror on one side and a massive input coupler
mirror on the other side (see figure \ref{setup}). The input-coupler
is a concave mirror with radius of curvature R = 25mm and a
reflectivity of 99.4\% for 1064nm radiation. The cavity was slightly
shorter than 25 mm. The chip containing the micro-mirror was placed
on a 3-axis translation stage for alignment
and the optical beam was positioned on the oscillator. The size of
the waist of the $TEM_{00}$ mode of the cavity is around 20
$\mu m$, which is much smaller than the width of the beam. The input
mirror is placed on a piezoelectric transducer (PTZ)respectively
 to scan the length of
the cavity. The whole cavity is placed in a vacuum chamber with a
pressure of $p < 2 \cdot 10^{-5}$ mbar to avoid damping of the
oscillations of the beam due to air friction. All experiments were
performed at room temperature.

For readout, the cavity is locked via the Pound-Drever-Hall (PDH)
method \cite{pdh}. The PDH locking signal is sent to a PID
controller, the output of which is sent to the PZT to control the
fine length of the cavity at resonance (Fig.3). It is known that the
intensity of the PDH error signal around resonance is proportional
to the change in length of the cavity. As a consequence, the
vibrational noise of the beam can be monitored with high accuracy
via the spectrum of the PDH error signal \cite{pdhmeasurement}. The
optical quality is characterized by measuring the finesse F of the
cavity, which is related to the round-trip intensity loss $\gamma$
via $F = 2 \pi / \gamma$ in the limit of a high-finesse cavity. We
measured a finesse of 650 at non-undercut regions of the mirror and
of around 500 on the beam, corresponding to overall round trip
losses of 1\% and 1.3\% respectively (This includes the
0.6\% insertion losses on the input coupler).
To determine the reflectivity of the coating, we build a cavity
using a region of the coating that is spaced a few mm from the
ablated structures.  In this regions we obtain a reflectivity of
99.6\% (i.e. a minimal degradation from the nominal value). The
additional losses of 0.3\%
on the beam can be explained both by diffraction losses and by
imperfections introduced during laser ablation, in particular near
the edges of the beam.

The mechanical quality factor is obtained via the frequency spectrum of the
PDH signal while the cavity is locked at resonance with the input
laser frequency. Several mechanical resonance peaks can be detected
within the range of 200 kHz to 5 MHz. Their frequencies match well
with finite elements (FE) simulations of transverse vibrational
modes (i.e. oscillations occur normal to the surface) of
a doubly clamped beam under tensile stress, specifically of a
fundamental mode at 278kHz and its harmonics. We observe an
additional splitting of each transverse mode into two modes
separated by a few kHz, which can be identified as modes with
different torsional contributions.

A critical parameter for performing high-sensitivity measurements
with movable micro-mirrors is their mechanical quality factor Q,
which is a direct measure for the time scale of dissipation in an
oscillating system. Q is defined as the ratio between the resonance
frequency and the full width at half maximum (FWHM)correspond to
 of the resonance
peak. It is a crucial parameter for sensors as it limits the
sensitivity of all measurements. We isolated the lowest resonance at
278 kHz, which corresponds to the fundamental transverse mode of the
beam. The measured Q-factor of this mode was around 9000, with a
FWHM of around 32Hz. The Q-factor of higher-order modes decreases
monotonically to around 2000 for the fourth transverse mode at
1.2MHz, which is indicative for the presence of clamping losses~(see
for example \cite{Ekinci_2005} and references therein). One should
finally note that our method is not limited to the present parameter
range but can in principle yield much higher reflectivity and Q. For
example, the reflectivity can be increased by employing
sophisticated state-of-the-art HR coatings as are typically used for
gravitational wave detectors (yielding reflectivity of up to
99.9999\%). Another possibility is to use Silicon-On-Insulator (SOI)
technology instead of plain silicon wafers. The dry etch would then
undercut only the device layer and stop on the buried oxide. This
could lead to more kinetically controlled etching, i.e. to a better
control of the membrane uniformity, and hence to a more controlled
resonance frequency and higher Q factor.

In conclusion, we have demonstrated a promising method for the
fabrication of micromechanical mirrors with high reflectivity, high
mechanical quality, and low mass. Such a Bragg mirror is the
lightest HR beam that one can design since the coating itself
constitutes all the mass. We have characterized its mechanical and
optical properties by using this micro-mirror in a Fabry-Perot
interferometer. The reflectivity of the mirror is in principle
limited only by the intrinsic coating quality. The combination of
high reflectivity, low mass and high mechanical quality make the
fabricated micro-mechanical mirrors an excellent candidate for
high-sensitivity measurements down to the quantum limit. In
addition, such structures may provide the possibility to study
genuine quantum effects involving mechanical
systems~\cite{Mancini94,Bose97,Marshall03,Pinard05}.

\begin{acknowledgments}
We would like to thank Heidi Piglmayer-Brezina for fabricating the
masks for the laser ablation. We acknowledge financial support by
the Austrian Science Fund (FWF) under the programs SFB15 and
P16133-N08, by the Austrian NANO Initiative (MNA), by the European
Commission under the Integrated Project Qubit Applications (QAP)
funded by the IST Directorate under contract number 015846, by the
Foundational Questions Institute (FQXI) and by the City of Vienna.
\end{acknowledgments}

\end{document}